%% file: DESY-07-012-PL.tex
\documentclass[zpreprint,zbstpl]{zeus_paper}
\usepackage[english]{babel}

\input zeus_def.tex

\def\citeCTD{{\cite{%
nim:a279:290,*npps:b32:181,*nim:a338:254%
}}\xspace}
\def\citeCAL{{\cite{%
nim:a309:77,*nim:a309:101,*nim:a321:356,*nim:a336:23%
}}\xspace}
\renewcommand{\Zctddesc}[1]{%
Charged particles are tracked in the central tracking detector (CTD)~\citeCTD,
which operates in a magnetic field of $1.43\Tesla$ provided by a thin 
superconducting coil. The CTD consists of 72~cylindrical drift chamber 
layers, organized in nine superlayers covering the polar-angle#1 region 
\mbox{$15^\circ<\theta<164^\circ$}. The transverse-momentum resolution for
full-length tracks is $\sigma(p_T)/p_T=0.0058p_T\oplus0.0065\oplus0.0014/p_T$,
with $p_T$ in $\Gev$.}
\renewcommand{\Zcaldesc}{%
The high-resolution uranium-scintillator calorimeter (CAL)~\citeCAL consists 
of three parts: the forward (FCAL), the barrel (BCAL) and the rear (RCAL)
calorimeters. Each part is subdivided transversely into towers and
longitudinally into one electromagnetic section and either one (in RCAL)
or two (in BCAL and FCAL) hadronic sections. The smallest subdivision of
the calorimeter is called a cell.  The CAL energy resolutions, as measured under
test-beam conditions, are $\sigma(E)/E=0.18/\sqrt{E}$ for electrons and
$\sigma(E)/E=0.35/\sqrt{E}$ for hadrons, with $E$ in $\Gev$.}

\newcommand{\dstar}     {\ensuremath{D^{\ast}}}
\newcommand{\dstarp}    {\ensuremath{D^{\ast +}}}
\newcommand{\dstarpm}   {\ensuremath{D^{\ast \pm}}}
\newcommand{\dzero}     {\ensuremath{D^{0}}}
\newcommand{\ptds}{\ensuremath{p_T(D^{\ast})}}
\newcommand{\etads}{\ensuremath{\eta(D^{\ast})}}
\newcommand{\siggp}{\ensuremath{\sigma_{\gamma{} p}}}
\newcommand{\qsq}{\ensuremath{{Q^{2}}}}
\newcommand{\pT}{\ensuremath{p_{T}}}
\newcommand{\ETtheta}{\ensuremath{E_{T}^{\theta>10^{\circ}}}}
\newcommand{\ftwo}      {\mbox{$F_{2}$}}

\newcommand{\pho}{\phantom{0}}

\begin{document}
\prepnum{DESY--07--012}

\title{
\boldmath Measurement of \dstarpm{} meson production\\ 
in $e^\pm p$ scattering at low \qsq 
}                                                       
                    
\author{ZEUS Collaboration}
\date{13th February 2007}

\abstract{The production of \dstarpm(2010) mesons in $e^\pm p$ scattering in
  the range of exchanged photon virtuality $0.05 < \qsq < 0.7\gev^{2}$ has
  been measured with the ZEUS detector at HERA using an integrated luminosity
  of 82 pb$^{-1}$.  The decay channels $\dstarp \rightarrow \dzero \pi^+$ with
  $\dzero \rightarrow K^-\pi^+$ and corresponding antiparticle decay were used
  to identify \dstar{} mesons and the ZEUS beampipe calorimeter was used to
  identify the scattered electron.  Differential \dstar{} cross sections as
  functions of \qsq, inelasticity, $y$, transverse momentum of the \dstar{}
  meson, \ptds, and pseudorapidity of the \dstar{} meson, \etads, have been
  measured in the kinematic region $0.02 < y < 0.85$, $1.5 < \ptds < 9.0\gev$
  and $|\etads| < 1.5$.  The measured differential cross sections are in
  agreement with two different NLO QCD calculations. The cross sections are
  also compared to previous ZEUS measurements in the photoproduction and DIS
  regimes.}

\makezeustitle

\def\3{\ss}                                                                                        
\pagenumbering{Roman}                                                                              
\begin{center}                                                                                     
{                      \Large  The ZEUS Collaboration              }                               
\end{center}                                                                                       
  S.~Chekanov$^{   1}$,                                                                            
  M.~Derrick,                                                                                      
  S.~Magill,                                                                                       
  S.~Miglioranzi$^{   2}$,                                                                         
  B.~Musgrave,                                                                                     
  D.~Nicholass$^{   2}$,                                                                           
  \mbox{J.~Repond},                                                                                
  R.~Yoshida\\                                                                                     
 {\it Argonne National Laboratory, Argonne, Illinois 60439-4815}, USA~$^{n}$                       
\par \filbreak                                                                                     
  M.C.K.~Mattingly \\                                                                              
 {\it Andrews University, Berrien Springs, Michigan 49104-0380}, USA                               
\par \filbreak                                                                                     
  M.~Jechow, N.~Pavel~$^{\dagger}$, A.G.~Yag\"ues Molina \\                                        
  {\it Institut f\"ur Physik der Humboldt-Universit\"at zu Berlin,                                 
           Berlin, Germany}
% ...
\newpage                                                                                           
$^{\    1}$ supported by DESY, Germany \\                                                          
$^{\    2}$ also affiliated with University College London, UK \\                                  
% ...
$^{\dagger}$ deceased \\                                                                           
%                                                                                                  
% \par         % if index listing & table fit to 1 page, put gap here                              
\newpage   % alternatively: go to newpage, if page is too small                                    
                                                           %                                       
% \institute_references_start    % do not touch or move this line !                                
                                                           %                                       
\begin{tabular}[h]{rp{14cm}}                                                                       
$^{a}$ &  supported by the Natural Sciences and Engineering Research Council of Canada (NSERC) \\  
$^{b}$ &  supported by the German Federal Ministry for Education and Research (BMBF), under        
          contract numbers HZ1GUA 2, HZ1GUB 0, HZ1PDA 5, HZ1VFA 5\\                                
$^{c}$ &  supported in part by the MINERVA Gesellschaft f\"ur Forschung GmbH, the Israel Science   
          Foundation (grant no. 293/02-11.2) and the U.S.-Israel Binational Science Foundation \\  
$^{d}$ &  supported by the German-Israeli Foundation and the Israel Science Foundation\\           
$^{e}$ &  supported by the Italian National Institute for Nuclear Physics (INFN) \\                
$^{f}$ &  supported by the Japanese Ministry of Education, Culture, Sports, Science and Technology 
          (MEXT) and its grants for Scientific Research\\                                          
$^{g}$ &  supported by the Korean Ministry of Education and Korea Science and Engineering          
          Foundation\\                                                                             
$^{h}$ &  supported by the Netherlands Foundation for Research on Matter (FOM)\\                   
$^{i}$ &  supported by the Polish State Committee for Scientific Research, grant no.               
          620/E-77/SPB/DESY/P-03/DZ 117/2003-2005 and grant no. 1P03B07427/2004-2006\\             
$^{j}$ &  partially supported by the German Federal Ministry for Education and Research (BMBF)\\   
$^{k}$ &  supported by RF Presidential grant N 8122.2006.2 for the leading                         
          scientific schools and by the Russian Ministry of Education and Science through its grant
          Research on High Energy Physics\\                                                        
$^{l}$ &  supported by the Spanish Ministry of Education and Science through funds provided by     
          CICYT\\                                                                                  
$^{m}$ &  supported by the Particle Physics and Astronomy Research Council, UK\\                   
$^{n}$ &  supported by the US Department of Energy\\                                               
$^{o}$ &  supported by the US National Science Foundation. Any opinion,                            
findings and conclusions or recommendations expressed in this material                             
are those of the authors and do not necessarily reflect the views of the                           
National Science Foundation.\\                                                                     
$^{p}$ &  supported by the Polish Ministry of Science and Higher Education\\                       
$^{q}$ &  supported by FNRS and its associated funds (IISN and FRIA) and by an Inter-University    
          Attraction Poles Programme subsidised by the Belgian Federal Science Policy Office\\     
$^{r}$ &  supported by the Malaysian Ministry of Science, Technology and                           
Innovation/Akademi Sains Malaysia grant SAGA 66-02-03-0048\\                                       
\end{tabular}                                                                                      
                                                           %                                       
% \institute_references_end     % do not touch or move this line !                                 
\clearpage
%------------------------------------------------------------------------------
%       Text
%------------------------------------------------------------------------------
\input{DESY-07-012-txt-PL.tex}
%------------------------------------------------------------------------------
%       Bibliography
%------------------------------------------------------------------------------
\clearpage
{\raggedright
\input{DESY-07-012-PL.bbl}

}
%------------------------------------------------------------------------------
%       Tables
%------------------------------------------------------------------------------
\input{DESY-07-012-tab.tex}
\clearpage
%------------------------------------------------------------------------------
%       Figures
%------------------------------------------------------------------------------
\input{DESY-07-012-fig.tex}
%
%       ... that's it
%
\end{document}

%% file: zeus_def.tex
\newcommand{\ZcoosysB}{%
The ZEUS coordinate system is a right-handed Cartesian system, with the $Z$
axis pointing in the proton beam direction, referred to as the ``forward
direction'', and the $X$ axis pointing left towards the centre of HERA.
The coordinate origin is at the nominal interaction point.\xspace}

\newcommand{\ZcoosysfnB}{\footnote{\ZcoosysB}}

\newcommand{\Zdetdesc}{%
A detailed description of the ZEUS detector can be found 
elsewhere~\cite{zeus:1993:bluebook}. A brief outline of the 
components that are most relevant for this analysis is given
below.\xspace}
\newcommand{\Zctddesc}[1]{%
Charged particles are tracked in the central tracking detector (CTD)~\citeCTD,
which operates in a magnetic field of $1.43\Tesla$ provided by a thin 
superconducting coil. The CTD consists of 72~cylindrical drift chamber 
layers, organized in 9~superlayers covering the polar-angle#1 region 
\mbox{$15^\circ<\theta<164^\circ$}. The transverse-momentum resolution for
full-length tracks is $\sigma(p_T)/p_T=0.0058p_T\oplus0.0065\oplus0.0014/p_T$,
with $p_T$ in $\Gev$.}
\newcommand{\Zcaldesc}{%
The high-resolution uranium--scintillator calorimeter (CAL)~\citeCAL consists 
of three parts: the forward (FCAL), the barrel (BCAL) and the rear (RCAL)
calorimeters. Each part is subdivided transversely into towers and
longitudinally into one electromagnetic section (EMC) and either one (in RCAL)
or two (in BCAL and FCAL) hadronic sections (HAC). The smallest subdivision of
the calorimeter is called a cell.  The CAL energy resolutions, as measured under
test-beam conditions, are $\sigma(E)/E=0.18/\sqrt{E}$ for electrons and
$\sigma(E)/E=0.35/\sqrt{E}$ for hadrons ($E$ in $\Gev$).}
\chardef\usc=95
\chardef\til=126
\catcode`\@=11 % @ signs are now treated as letters
\DeclareRobustCommand\xdotspace{\futurelet\@let@token\@xdotspace}
\def\@xdotspace{%
  \ifx\@let@token.\else
  \ifx\@let@token\bgroup.\else
  \ifx\@let@token\egroup.\else
  \ifx\@let@token\/.\else
  \ifx\@let@token\ .\else
  \ifx\@let@token~.\else
  \ifx\@let@token!.\else
  \ifx\@let@token,.\else
  \ifx\@let@token:.\else
  \ifx\@let@token;.\else
  \ifx\@let@token?.\else
  \ifx\@let@token/.\else
  \ifx\@let@token'.\else
  \ifx\@let@token).\else
  \ifx\@let@token-.\else
  \ifx\@let@token\@xobeysp.\else
  \ifx\@let@token\space.\else
  \ifx\@let@token\@sptoken.\else
   .\space
   \fi\fi\fi\fi\fi\fi\fi\fi\fi\fi\fi\fi\fi\fi\fi\fi\fi\fi}
\catcode`\@=12 % @ signs are no longer letters

\newcommand{\stru}[2]{%
   \relax\ifmmode\hbox{\vrule height#1 depth#2 width0pt}%
   \else\vrule height#1 depth#2 width0pt\fi}
\newcommand{\Ronum}[1]{\uppercase\expandafter{\romannumeral#1}}
\newcommand{\ronum}[1]{\expandafter{\romannumeral#1}}
\DeclareRobustCommand{\LaTeXZ}{%
  \LaTeX\kern-.05em4\kern-.1em
  {\raisebox{-0.2ex}{$\scriptstyle\text{ZEUS}$}}\xspace}
\DeclareMathAlphabet{\mathbf}{OT1}{cmr}{bx}{sl}
\newcommand{\eVdist}{\kern-0.06667em}

\newcommand{\Gev}{{\text{Ge}\eVdist\text{V\/}}}

\newcommand{\gev}{{\,\text{Ge}\eVdist\text{V\/}}}

\newcommand{\nb}{\,\text{nb}}
\newcommand{\pb}{\,\text{pb}}

\newcommand{\cm}{\,\text{cm}}

\newcommand{\Tesla}{\,\text{T}}

\newcommand{\slashfrac}[2]{%
  \raisebox{0.5ex}{\ensuremath #1}\kern-0.12em/\kern-0.08em
  \raisebox{-.8ex}{\ensuremath #2}}
\newcommand{\sqr}[3]{%
    {\vcenter{\hrule height.#3ex\hbox{\vrule width.#2ex height#1ex
     \kern#1ex\vrule width.#3ex}\hrule height.#2ex}}}

\catcode`\@=11 % @ signs are now treated as letters
\newcommand{\parenbar}{\mathpalette\p@renb@r}
\def\p@renb@r#1#2{\vbox{%
  \ifx#1\scriptscriptstyle \dimen@.7em\dimen@ii.2em\else
  \ifx#1\scriptstyle \dimen@.8em\dimen@ii.25em\else
  \dimen@1em\dimen@ii.4em\fi\fi \offinterlineskip
  \ialign{\hfill##\hfill\cr
    \vbox{\hrule width\dimen@ii}\cr
    \noalign{\vskip-.3ex}%
    \hbox to\dimen@{$\mathchar300\hfil\mathchar301$}\cr
    \noalign{\vskip-.3ex}%
    $#1#2$\cr}}}
\catcode`\@=12 % @ signs are no longer letters

\newcommand{\rnge}{\hbox{$\,\text{--}\,$}}

\newcommand{\IP}{{\rm I$\kern-0.01667em$P}\xspace}

\mathchardef\qsm=63
\mathchardef\pls=43
\mathchardef\mns=512
\mathchardef\plm=518
\mathchardef\eql=61
\mathchardef\smallleft=300
\mathchardef\smallright=301
\mathchardef\les=316
\mathchardef\gre=318
\mathchardef\leq=532
\mathchardef\grq=533
\catcode`\@=11 % @ signs are now treated as letters
\newcounter{pict@width}
\newcounter{pict@height}
\newlength{\pict@scale}
\newcommand{\psfigadd}[4]{%
\setcounter{pict@width}{1*\ratio{#2+\pict@scale/2}{\pict@scale}}
\setcounter{pict@height}{1*\ratio{#3+\pict@scale/2}{\pict@scale}}
\setlength{\unitlength}{\pict@scale}
\hbox to #2{\hspace{-\fill}\begin{picture}(\thepict@width,\thepict@height)
\put(0,0){\psfig{figure=#1,width=#2,height=#3,clip=}}
\SetScale{0.283466457}
\SetWidth{1.763889}
{#4}
\end{picture}}
}
\newcounter{pict@widthfst}
\newcounter{pict@widthscd}
\newcounter{pict@widthtot}
\newcommand{\psfigaddtwo}[7]{%
\setcounter{pict@widthfst}{1*\ratio{#2+\pict@scale/2}{\pict@scale}}
\setcounter{pict@widthscd}{1*\ratio{#2+#4+\pict@scale/2}{\pict@scale}}
\setcounter{pict@widthtot}{1*\ratio{#2+#4+#6+\pict@scale/2}{\pict@scale}}
\setcounter{pict@height}{1*\ratio{#3+\pict@scale/2}{\pict@scale}}
\setlength{\unitlength}{\pict@scale}
\hbox{\hspace{-\fill}\begin{picture}(\thepict@widthtot,\thepict@height)
\put(0,0){\psfig{figure=#1,width=#2,height=#3,clip=}}
\put(\thepict@widthscd,0){\psfig{figure=#5,width=#6,height=#3,clip=}}
\SetScale{0.283466457}
\SetWidth{1.763889}
{#7}
\end{picture}}
}
\newcommand{\psfigror}[4]{%
\setcounter{pict@width}{1*\ratio{#2+\pict@scale/2}{\pict@scale}}
\setcounter{pict@height}{1*\ratio{#3+\pict@scale/2}{\pict@scale}}
\setlength{\unitlength}{\pict@scale}
\hbox{\begin{picture}(\thepict@width,\thepict@height)
\put(0,\thepict@height){\psfig{figure=#1,width=#3,height=#2,clip=,angle=270}}
\SetScale{0.283466457}
\SetWidth{1.763889}
{#4}
\end{picture}}
}
\newcommand{\psfigrol}[4]{%
\setcounter{pict@width}{1*\ratio{#2+\pict@scale/2}{\pict@scale}}
\setcounter{pict@height}{1*\ratio{#3+\pict@scale/2}{\pict@scale}}
\setlength{\unitlength}{\pict@scale}
\hbox{\begin{picture}(\thepict@width,\thepict@height)
\put(0,0){\psfig{figure=#1,width=#3,height=#2,clip=,angle=90}}
\SetScale{0.283466457}
\SetWidth{1.763889}
{#4}
\end{picture}}
}
\catcode`\@=12 % @ signs are no longer letters
\newlength\listtextwidth

\catcode`\@=11 % @ signs are now treated as letters
\newlength{\@tabfninsert}
\newlength{\@tabfnwidth}
\newcommand{\tabfootnote}[2]{%
  \setlength{\@tabfninsert}{0.8em}
  \setlength{\@tabfnwidth}{\textwidth}
  \addtolength{\@tabfnwidth}{-\@tabfninsert}
  \addtolength{\@tabfnwidth}{-0.4em}
  \noindent\makebox[\@tabfninsert][r]{\footnotesize$^{#1}$\hfil}\hfill%
  \parbox[t]{\@tabfnwidth}{\footnotesize #2\hfill}}
\catcode`\@=12 % @ signs are no longer letters

%% file: DESY-07-012-txt-PL.tex
\pagenumbering{arabic} 
\pagestyle{plain}
%
% ----------------------------------------------------------------------------
% Introduction
% ----------------------------------------------------------------------------
%
\section{Introduction}
\label{sec:int}
The production of charm quarks at HERA has been studied both in deep inelastic
scattering
(DIS)~\cite{pl:b407:402,np:b545:21,epj:c12:35,pl:b528:199,pr:d69:012004} and
photoproduction~\cite{np:b472:32,epj:c6:67,np:b729:492,epj:c47:597,hep-ex-0608042}.  
In general, reasonable agreement
is seen with next-to-leading-order (NLO) QCD predictions.

This paper presents measurements of the \dstar{} cross section in the range
$0.05 < \qsq < 0.7\gev^{2}$.
The beampipe calorimeter of
ZEUS~\cite{pl:b407:432,pl:b487:53} was used for the measurement of the
scattered lepton, which allows the first measurements of the transition region
between photoproduction (photon virtuality, $\qsq \sim 0\gev^2$) and DIS
($\qsq > 1\gev^{2}$).
The cross sections are compared to the predictions of two different NLO QCD
calculations, one designed for DIS, the other for the photoproduction region.
This paper investigates whether the calculations remain valid in this
transition region.

%
% ----------------------------------------------------------------------------
% Experimental set-up
% ----------------------------------------------------------------------------
%
\section{Experimental set-up}
\label{sec:exp}
This analysis was performed with data taken from 1998 to 2000, when HERA
collided electrons or positrons\footnote{Hereafter, both electrons and
  positrons are referred to as electrons.}  with energy $E_e=27.5\gev$ with
protons of energy $E_p=920\gev$.  The combined data sample has an integrated
luminosity of $\mathcal{L}=81.9 \pm 1.8$\,pb$^{-1}$.

\Zdetdesc

\Zctddesc\ZcoosysfnB

\Zcaldesc

The scattered electron was detected in the beampipe
calorimeter (BPC).  The BPC 
allowed the detection of low-\qsq{} events, where the electron is scattered
through a small angle. The BPC was used in previous measurements of the proton
structure function, $\ftwo$, at low \qsq~\cite{pl:b407:432,pl:b487:53}.  It
originally consisted of two tungsten--scintillator sampling calorimeters with
the front faces located at $Z=-293.7\cm$, the centre at $Y=0.0\cm$, and the
inner edge of the active area at $X=\pm 4.4\cm$, as close as possible to the
electron-beam trajectory. 
At the end of 1997 one of the two BPC calorimeters was removed; hence, for the
analysis in this paper, only the calorimeter located on the $+X$ side of the
beampipe was utilised.
It had an active area of $12.0 \times
12.8\cm^2$ in $X\times Y$ and a depth of $24$ radiation lengths. The relative
energy resolution as determined in test-beam measurements with 1\rnge{}6\gev{}
electrons was ${\Delta E}/{E}={17\%}/{\sqrt{E\,(\gev)}}$.

The luminosity was measured from the rate of the bremsstrahlung process $ep
\rightarrow e\gamma p$, where the photon was measured in a lead--scintillator
calorimeter\cite{desy-92-066,*zfp:c63:391,*acpp:b32:2025} placed in the HERA
tunnel at Z=-107m.

A three-level trigger system was used to select events
online\cite{zeus:1993:bluebook,proc:chep:1992:222}.  At all three
levels, the event was required to contain a scattered electron candidate in
the BPC.  Additionally, at the third level, a reconstructed \dstar{} candidate
was required for the event to be kept for further analysis.  The efficiency of
the online \dstar{} reconstruction, determined relative to an inclusive
DIS trigger, was above 95$\%$\cite{pr:d69:012004}.

%
%--------------------------------------------------------------------
% Kinematics
%--------------------------------------------------------------------
%
\section{Kinematic reconstruction and event selection}
\label{sec:kinvar}

Deep inelastic electron-proton scattering, $ep\to eX$, can be described in
terms of two kinematic variables, chosen here to be $y$ and \qsq, where $y$ is
the inelasticity.
They are defined as $\qsq=-q^2=-(k-k')^2$ and $y=\qsq/(2P\cdot q)$,
where $k$ and $P$ are the four-momenta of the incoming electron and proton,
respectively, and $k'$ is the four-momentum of the scattered electron. The
inelasticity, which is the fractional energy transferred to the proton in its
rest frame, is related to the Bjorken scaling variable $x$ and \qsq{} by
$\qsq=sxy$, where $s=4E_{e}E_{p}$ is the square of the electron-proton
centre-of-mass energy of 318\gev.

The values of $y$ and \qsq{} were calculated using the measured electron
scattering angle and the energy deposited in the BPC as detailed in a previous
analysis\cite{pl:b407:432}, which also describes the method used for the
energy calibration of the BPC.  A time dependent re-calibration of the energy
response was necessary~\cite{thesis:tandler:2003}, 
as radiation damage of the
scintillator resulted in a degradation of about $10\%$ by the end of the 2000
running period.

A series of cuts was applied
to reject background.  The events were required to have a primary vertex
within 50\cm{} in $Z$ of the nominal interaction point.  The electron
candidates in the BPC were required to have $E_{\mathrm{BPC}}>4\gev$, as the
trigger efficiency is low below this energy.  The electron impact point on the
face of the BPC was required to be more than 0.7\cm{} from the inner edge to
ensure good shower containment.
Photoproduction events were efficiently rejected by requiring the events to
have $35 < E-P_{Z} < 65 \gev$, where $E-P_{Z} = \sum_{i} (E-P_{Z})_{i}$ is
summed over all CAL deposits, including the scattered electron candidate in
the BPC.  Finally, events with an additional well-reconstructed electron
candidate in the CAL with energy greater than 5\gev{} were rejected to reduce
background from DIS events with $\qsq > 1\gev^{2}$.

The measured kinematic region in $y$ and $\qsq$ was restricted to the range of
high acceptance, $0.02 < y < 0.85$, $0.05 < \qsq{} < 0.7\gev^2$.  With these
cuts, the reconstructed invariant mass of the hadronic system, $W$, lies
between 50 and 300\gev, with a 
mean of 190\gev.

%
% ----------------------------------------------------------------------------
% Selection of the D* candidates
% ----------------------------------------------------------------------------
%
\section{\boldmath Selection of \dstar{} candidates}
\label{sub:dssel}

The \dstar{} mesons were identified using the decay channel $\dstarp \to \dzero
\pi^{+}_s$ with the subsequent decay $\dzero \to K^-\pi^+$ and the
corresponding antiparticle decay chain, where $\pi^+_s$ refers to a
low-momentum (``slow'') pion accompanying the \dzero.

Charged tracks measured by the CTD and assigned to the primary event
vertex\footnote{The resolution of such tracks is not good enough to separate
  primary and secondary vertices from $c$ and $b$ hadron decays.}
were selected. The transverse momentum was required to be greater than
0.12\gev. The \pT{} cut was raised to 0.25\gev{} for a data subsample
corresponding to $(16.9 \pm 0.4)\pb^{-1}$, for which  the
low-momentum track-reconstruction efficiency was lower due to the operating
conditions of the CTD\cite{nim:a515:37}.  Each track was required
to reach at least the third superlayer of the CTD. These restrictions ensured
that the track acceptance was high and the momentum resolution was good.
Tracks in the CTD with opposite charges and transverse momenta $p_T >
0.45\gev$ were combined in pairs to form $D^0$ candidates. The tracks were
alternately assigned the kaon and the pion mass and the invariant mass of
the pair, $M_{K\pi}$, was determined. Each additional track, with charge
opposite to that of the kaon track, was assigned the pion mass and combined
with the $D^0$-meson candidate to form a \dstar{} candidate.

A mass window for the signal region of the \dzero{} varying from $1.82<
M_{K\pi} < 1.91\gev$ to $1.79< M_{K\pi} < 1.94\gev$ was used, 
reflecting the dependence of the CTD resolution on $\ptds$.  The signal
region for the reconstructed mass difference $\Delta M=(M_{K\pi\pi_s} -
M_{K\pi})$ was \mbox{$0.1435 < \Delta M < 0.1475\gev$}.  The requirement of
$\ptds/\ETtheta > 0.1$ was also applied, where \ETtheta{} is the transverse
energy outside a cone of $\theta = 10^{\circ}$ defined with respect to the
proton direction. This cut rejects background without significantly
affecting the signal.

The \dstar{} mesons were selected in the kinematic region $1.5<\ptds<9\gev$
and $|\etads|<1.5$.  The $\Delta M$ distribution for events with an
electron reconstructed in the BPC is shown in Fig.~\ref{fig1}.  
To extract
the number of \dstar{} mesons, the $\Delta M$ distribution was fit using an
unbinned likelihood method, with a Gaussian to describe the signal and a
threshold function to describe the combinatorial background.  
A first estimate of the background was given by \dstar{} candidates with
wrong-sign combinations, in which both tracks forming the \dzero{} candidates
have the same charge and the third track has the opposite charge. 
These are
shown as the shaded region in Fig.~\ref{fig1}.
The number of \dstar{} mesons obtained from the fit was $N(\dstar)=253 \pm 25$.

%
% ----------------------------------------------------------------------------
%       Event simulation and systematics
% ----------------------------------------------------------------------------
%
\section{Acceptance corrections and systematic uncertainties}
\label{sec:evsim}
  
The acceptances were calculated using the {\sc
Herwig~6.1}~\cite{hep-ph-9912396,*cpc:67:465} and {\sc
Rapgap~2.08}~\cite{cpc:86:147} Monte Carlo (MC) models. 
Both models simulate charm and beauty production and include contributions
from both direct and resolved photoproduction.
In direct photoproduction the photon participates as a point-like particle in
the hard scattering process,
while in resolved photoproduction a parton in the photon
scatters on a parton in the proton.
The generated events were passed
through a full simulation of the detector,
using {\sc Geant 3.13}~\cite{tech:cern-dd-ee-84-1} and then processed and
selected with the same programs as used for the data.  The
CTEQ5L~\cite{epj:c12:375} parton density function (PDF) was used for the
proton and GRV-LO~\cite{pr:d46:1973} was used for the photon. The charm-quark
mass was set to 1.5\gev.

The {\sc Herwig} predictions are in good agreement with the data distributions
for both the scattered lepton and hadronic variables and so this Monte Carlo
was used to correct the data for detector effects.  For the kinematic region
of the measurement 
$0.05<\qsq<0.7\gev^2$, $0.02<y<0.85$, $1.5<\ptds<9\gev$, and
$|\etads|<1.5$ the acceptance was $(1.11\pm 0.03)\%$. This includes
the geometrical acceptance of the BPC, which was about 9\%, and the
reconstruction efficiency for the \dstar{} decay chain.

The {\sc Rapgap} MC gives a similarly good representation of the 
data and was used to estimate part of the systematic 
uncertainties, as described below.

The differential cross section for a given observable $Y$ was determined
using
\begin{eqnarray*}
\frac {d\sigma}{dY} & = &
\frac {N} {A \cdot \mathcal {L} \cdot B \cdot \Delta Y},  
\end{eqnarray*}

where $N$ is the number of \dstar{} events in a bin of size $\Delta Y$, $A$ is
the acceptance (which takes into account migrations and efficiencies for that
bin) and $\mathcal {L}$ is the integrated luminosity. The product, $B$, of the
appropriate branching ratios for the \dstar{} and \dzero{} decays was set to
$(2.57\pm 0.05)\%$~\cite{jphys:g33:1}.

The systematic uncertainties of the measured cross sections were determined by
changing in turn the selection cuts or the analysis procedure within their
uncertainties and repeating the extraction of the cross
sections~\cite{thesis:irrgang:2004}.  The major experimental sources of
systematic uncertainty were (the variation of the total cross section is given
in parentheses): the BPC alignment ($^{+2.5}_{-3.1} \%$) and energy scale
($^{+0.4}_{-1.2} \%$); the uncertainty in the CTD momentum scale
($^{+0.2}_{-1.5} \%$) and the CAL energy scale ($\pm 1 \%$); the
$\ptds/\ETtheta$ cut ($^{+3.0}_{-1.7} \%$) and the \dstar{} signal extraction
($^{+0.1}_{-1.5} \%$).  The uncertainty due to the MC model ($^{+9.5}_{-4.8}
\%$) was determined by using {\sc Rapgap} to evaluate the acceptance
correction rather than {\sc Herwig}, as well as by varying the fraction of
resolved and direct photoproduction processes in the simulation.

All the above errors were added in quadrature separately for the positive and
negative variations to determine the overall systematic uncertainty.  The
overall normalisation has additional uncertainties of 2.2\% due to the
luminosity measurement and 2.0\%
due to knowledge of branching ratios. These are
included in the error quoted for the total cross section but not in
the systematic uncertainties of the differential cross sections.

%
% ----------------------------------------------------------------------------
%       Theoretical predictions
% ----------------------------------------------------------------------------
%
\section{Theoretical predictions}
\label{sec:theory}

Two different calculations were used to
evaluate the theoretical expectation for charm production.

The HVQDIS program~\cite{pr:d57:2806} implements an NLO calculation of charm
production in DIS. At low \qsq{}, the hadron-like structure of the photon, not
included in HVQDIS, is needed to regularise the NLO calculation. Therefore
predictions from this program are expected to lose accuracy in the 
limit $\qsq{} \rightarrow 0$. The ZEUS measurements of \dstar{} production in
DIS for 
$\qsq > 1.5\gev^{2}$ are in good agreement with the HVQDIS
prediction~\cite{pr:d69:012004}. 

The FMNR program~\cite{np:b454:3} implements an NLO calculation of charm
photoproduction which includes the hadron-like component of the photon.
Electroproduction cross sections can be obtained with FMNR using the
Weizs\"acker-Williams approximation~\cite{pl:b319:339} and are therefore
expected to be reliable only at low \qsq, where this approximation is valid.
The FMNR predictions are in reasonable agreement with ZEUS
measurements of \dstar{} photoproduction~\cite{epj:c6:67}, considering the
theoretical uncertainties.

It is therefore interesting to see whether these calculations are able to
reproduce the data in the transition region between photoproduction and DIS.
The following parameters were used in the calculations for both programs.
They were chosen to be the same as in a previous
publication~\cite{pr:d69:012004}. A variant of the ZEUS-S NLO QCD global
fit~\cite{pr:d67:012007} to structure-function data was used as the
parameterisation of the proton PDFs. This fit was repeated in the
fixed-flavour-number scheme, FFNS, in which the PDF has three active quark
flavours in the proton, and $\Lambda^{(3)}_{\rm QCD}$ is set to 0.363\gev. The
mass of the charm quark was set to 1.35\gev.  The renormalisation and
factorisation scales were set to $\mu_{R} = \mu_{F} = \sqrt{\qsq+4m_c^2}$ in
HVQDIS, while for FMNR they were set to the usual choice of $\mu_{R} = \mu_{F}
= \sqrt{\pT^2+m_c^2}$, where $\pT^2$ is the average transverse momentum
squared of the charm quarks. The charm fragmentation to a \dstar{} is carried
out using the Peterson function~\cite{pr:d27:105}. The hadronisation fraction,
$f(c \to \dstar)$, was taken to be $0.238$~\cite{hep-ex-9912064,*epj:c44:351}
and the Peterson parameter, $\epsilon$, was set to 0.035\cite{np:b565:245}.
The parameters used here for the FMNR calculation are different from those
used in a previous photoproduction analysis~\cite{epj:c6:67} (which used $m_c
= 1.5\gev$) leading to a 20\% larger predicted photoproduction cross section.

For the FMNR calculation the electroproduction cross section, $\sigma_{ep}$, was
obtained from the photoproduction cross section, $\sigma_{\gamma p}(W)$, using
\begin{eqnarray*}
  \sigma_{ep} & = & \int_{y_{\mathrm{min}}}^{y_\mathrm{max}} dy\,
    \Phi(y,Q^{2}_{\mathrm{min}},Q^{2}_{\mathrm{max}}) 
    \sigma_{\gamma p}(\sqrt{y s}),
\end{eqnarray*}
where
\begin{eqnarray}
  \label{eq:flux}
  \Phi(y,Q^{2}_{\mathrm{min}},Q^{2}_{\mathrm{max}}) & = &
  \frac{\alpha_{\mathrm{em}}}{2\pi} \left[
    \frac{(1 + (1-y)^{2})}{y} 
    \ln{\frac{Q^{2}_{\mathrm{max}}}{Q^{2}_{\mathrm{min}}}} -
    2 m_{e} y \left(
      \frac{1}{Q^{2}_{\mathrm{min}}} -
      \frac{1}{Q^{2}_{\mathrm{max}}}
    \right)
  \right]
\end{eqnarray}
is the photon flux and $y_{\mathrm{min}}$, $y_{\mathrm{max}}$, 
$Q^{2}_{\mathrm{min}}$, $Q^{2}_{\mathrm{max}}$ define the measurement range in
$y$ and $Q^{2}$.

The NLO QCD predictions for \dstar{} production are affected by systematic
uncertainties, which were also evaluated as in a previous ZEUS
paper~\cite{pr:d69:012004}\footnote{For the HVQDIS case, following
  \cite{pr:d69:012004}, the minimum value for the scales was 
  set to $2 m_c$.}.
The sources of systematic uncertainties on the total cross section are:
charm quark mass ($^{+15}_{-13}\%$ for HVQDIS,
 $^{+16}_{-14}\%$ for FMNR);
renormalisation and factorisation scale 
($^{+1}_{-13}\%$ for HVQDIS,
$^{+23}_{-10}\%$ for FMNR);
ZEUS PDF ($\pm 5\%$);
fragmentation ($^{+10}_{-6}\%$).
For both programs, the systematic uncertainties were added in quadrature and
are displayed as a band in the figures.  

Theoretical calculations of the total charm cross section in this
\qsq{} range can not be compared to the present data since
\dstar{} are only measured in a limited \pT{} and $\eta$ range.

%
% ----------------------------------------------------------------------------
%       Cross Sections
% ----------------------------------------------------------------------------
%
\section{Cross section measurements}
\label{sec:res}

The total cross section for $0.05<\qsq<0.7\gev^2$, $0.02<y<0.85$, 
\mbox{$1.5<\ptds<9\gev$} and $|\etads|<1.5$ is:
$$
\sigma(ep \rightarrow e \dstar{} X) =
10.1 \pm 1.0(\mbox{stat.})^{+1.1}_{-0.8}(\mbox{syst.})  \pm 0.20(\mbox{BR}) \nb,
$$
where the first uncertainty is statistical, the second from systematic effects
(including the luminosity uncertainty) and the third from the uncertainties in
the branching ratios.

The prediction from the HVQDIS program is $8.6^{+1.9}_{-1.8}\nb$, in
agreement with the data,  
while the prediction from FMNR is
$8.9^{+2.4}_{-1.4}\nb$\footnote{The contribution from the hadron-like
  component of the photon is 9\%.}, also in good agreement.

The measured differential \dstar{} cross sections as a function of \qsq, $y$,
\ptds{} and \etads{} for the data are shown in
Fig.~\ref{fig:hvqq2yetapt} and given in Table~\ref{tab:dxs}.
The predictions of the NLO calculations, including their uncertainties, are
shown as bands.
The measured differential cross sections 
are well described over the full measured kinematic region by both
calculations. 

This analysis was also compared to previous ZEUS measurements of \dstar{}
production in DIS~\cite{pr:d69:012004} made in the kinematic region
$1.5<\qsq<1000\gev^2$, $0.02<y<0.7$, $1.5<\ptds<15\gev$ and $|\etads|<1.5$.
In order to directly compare with the results presented there, the cross
sections were recalculated in the modified kinematic region $0.02<y<0.7$. No
correction was made for the different upper cut on \ptds{}, as the size of the
effect is $\approx 1\%$.

For this modified kinematic region, the differential cross section as a
function of \qsq{} is presented in Fig.~\ref{fig:dis2003ext} and given in
Table~\ref{tab:dxsr}. The systematic errors were assumed to be the same as
those in the full $y$ range. Figure~\ref{fig:dis2003ext} also shows the
previous ZEUS measurement and the HVQDIS prediction.  The combination of both
measurements shows that the slope of $d\sigma/d\qsq$ changes with \qsq; at
high \qsq{} the slope is steeper than at low \qsq.  
The NLO calculation describes the measured data well over the 
full \qsq{} range.

The \dstar{} electroproduction cross sections were converted to
$\gamma p$ cross sections, \siggp, in the range $1.5 < \ptds < 9\gev$ and
$|\etads| < 1.5$ (measured in the laboratory frame) using the photon flux from
Eq.~\ref{eq:flux}.  The cross sections are given for $W = 160\gev$, which
corresponds to $y=0.25$, close to the mean $y$ of the measured cross sections.
The $W$ dependence of \siggp{} was evaluated from the data. The uncertainty of
this procedure was estimated to be 10\%.  A comparison of the charm
photoproduction cross section~\cite{epj:c6:67}, this measurement and the DIS
cross sections~\cite{pr:d69:012004} is shown in Fig.~\ref{fig:gammapxsect}.
The numbers are tabulated in Table~\ref{tab:gammapxsect}.  The photoproduction
point was corrected for the different kinematic range and centre-of-mass
energy used here using the FMNR
program.
As can be seen,
the present measurements are consistent with the photoproduction cross
section.  A fit using a function of the form 
$\sigma(\qsq) = S M^{2} / (\qsq + M^2)$, 
where $S$ is the photoproduction cross section at $\qsq = 0$ and $M^2$
is the scale at which the $\gamma p$ cross section changes from the
photoproduction value to the DIS $1/\qsq$ behaviour, gives a good description
of the data over the whole \qsq{} range with $S = 823 \pm 63\nb$ and $M^{2} =
13 \pm 2\gev^{2}$. The value of $M^{2}$ found here for charm production is
close to $4 m_{c}^{2}$~\cite{prep:15:181} and significantly larger than that 
found for inclusive data $M_{0}^{2} = 0.52 \pm 0.05\gev^{2}$~\cite{pl:b487:53}.

%
% ----------------------------------------------------------------------------
%       Conclusions
% ----------------------------------------------------------------------------
%
\section{Conclusions}
\label{sec:concl}

Charm production has been measured as a function of \qsq, $y$, $\ptds$
and $\etads$ in the kinematic region $0.05 < \qsq < 0.7\gev^2$, $0.02 < y <
0.85$, $1.5 < \ptds < 9.0\gev$ and $|\etads | < 1.5$.  These measurements
extend the previous ZEUS measurements in DIS to lower \qsq.  The measured
differential cross sections are well described by two different NLO QCD
calculations: one (FMNR) is designed for the photoproduction region;
while the other (HVQDIS) is designed for DIS. Both calculations
predict similar cross sections in the intermediate \qsq{} region measured
here, which agree well with the measurements. 
The measurements, converted to $\gamma p$ cross sections, also agree well
with the \dstar{} photoproduction data.

%
% ----------------------------------------------------------------------------
%       Acknowledgements
% ----------------------------------------------------------------------------
%
\section*{Acknowledgements}

We would like to thank B.~Harris, E.~Laenen and S.~Frixione for helpful
discussions on the application of QCD calculations in this intermediate
regime.  We thank the DESY Directorate for their strong support and
encouragement. The remarkable achievements of the HERA machine group were
essential for the successful completion of this work.  The design,
construction and installation of the ZEUS detector have been made possible by
the effort of many people who are not listed as authors.

%%% Local Variables: 
%%% mode: latex
%%% TeX-master: "paper"
%%% End: 

%% file: DESY-07-012-PL.bbl
\providecommand{\etal}{et al.\xspace}
\providecommand{\coll}{Collaboration}
\catcode`\@=11
\def\@bibitem#1{%
\ifmc@bstsupport
  \mc@iftail{#1}%
    {;\newline\ignorespaces}%
    {\ifmc@first\else.\fi\orig@bibitem{#1}}
  \mc@firstfalse
\else
  \mc@iftail{#1}%
    {\ignorespaces}%
    {\orig@bibitem{#1}}%
\fi}%
\catcode`\@=12

%% file: DESY-07-012-tab.tex
\begin{table}[p]
  \centering
  \begin{tabular}{|c|ccc|}
    \hline 
    $Q^2$ bin & $d\sigma / dQ^2$ & $\Delta_{\rm stat}$ & $\Delta_{\rm syst}$ \\ 
    ($\gev^2$) & \multicolumn{3}{c|}{($\nb/\gev^{2}$)} \\
    \hline
    0.05:0.20    & $29.1$       & $\pm 7.2$   &   $ ^{+4.3} _{-4.1}$\\
    0.20:0.35    & $15.0$       & $\pm 2.4$   &   $ ^{+1.5} _{-1.4}$\\
    0.35:0.50    & $10.7$       & $\pm 2.2$   &   $ ^{+1.3} _{-1.1}$\\
    0.50:0.70    & $\pho 7.1$   & $\pm 2.3$   &   $ ^{+1.6} _{-0.8}$\\
    \hline 
    \hline
    $y$ bin & $d\sigma / dy$ & $\Delta_{\rm stat}$ & $\Delta_{\rm syst}$ \\
    & \multicolumn{3}{c|}{(\nb)} \\
    \hline
    0.02:0.15 & $34.2$ & $\pm 6.7$  & $ ^{+7.5} _{-7.5}$\\
    0.15:0.30 & $19.5$  & $\pm 3.8$ & $ ^{+2.7} _{-2.1}$\\
    0.30:0.50 & $10.7$   & $\pm 2.1$ & $ ^{+1.2} _{-1.1}$\\
    0.50:0.85 & $\pho 3.8$  & $\pm 1.1$  & $ ^{+0.8} _{-0.8}$ \\
    \hline
    \hline
    $p_T(D^*)$ bin & $d\sigma / dp_T(D^*)$ & $\Delta_{\rm stat}$ & $\Delta_{\rm syst}$  \\
    ($\gev$) & \multicolumn{3}{c|}{($\nb/\gev$)} \\
    \hline
    1.5:2.5  & $6.8$    & $\pm 1.7$    & $ ^{+1.0} _{-0.9}$   \\
    2.5:3.8  & $2.2$    & $\pm 0.3$    & $ ^{+0.2} _{-0.2}$ \\
    3.8:5.0  & $0.53$   & $\pm 0.11$   & $ ^{+0.02} _{-0.02}$ \\
    5.0:9.0  & $0.11$   & $\pm 0.02$   & $ ^{+0.01} _{-0.01}$   \\
    \hline
    \hline
    $\eta(D^*)$ bin & $d\sigma / d\eta(D^*)$ & $\Delta_{\rm stat}$ & $\Delta_{\rm syst}$  \\
    & \multicolumn{3}{c|}{(\nb)} \\
    \hline
    -1.5: -0.5  & $3.4$ & $ \pm 0.6$ & $ ^{+0.7} _{-0.7}$ \\
    -0.5:  0.0  & $4.1$ & $ \pm 0.9$ & $ ^{+0.5} _{-0.4}$ \\
    0.0:   0.5  & $2.9$ & $ \pm 0.8$ & $ ^{+0.3} _{-0.3}$ \\
    0.5:   1.5  & $3.3$ & $ \pm 0.7$ & $ ^{+0.4} _{-0.3}$ \\
    \hline
  \end{tabular}
  \caption{\textit{%
      Measured differential cross sections as a function of 
      \qsq, $y$, \ptds{} and \etads{} for $0.05<\qsq<0.7\gev^{2}$, $0.02<y<0.85$, 
      $1.5 < \ptds < 9\gev$ and $|\etads|<1.5$. The statistical and systematic 
      uncertainties are shown separately.
      The normalisation uncertainties from the luminosity measurement 
      and the branching ratios are not included in the systematic uncertainties.
    }}
  \label{tab:dxs}
\end{table}

\clearpage 

\begin{table}[p]
  \centering
  \begin{tabular}{|c|cc|}
    \hline 
    $\qsq$ bin & $d\sigma / d\qsq$ & $\Delta_{\rm stat}$ \\
    ($\gev^2$) & \multicolumn{2}{c|}{($\nb/\gev^{2}$)} \\
    \hline
    0.05:0.20    & $30.0$   & $\pm 7.2$    \\
    0.20:0.35    & $14.0$   & $\pm 2.3$    \\
    0.35:0.50    & $10.3$   & $\pm 2.1$    \\
    0.50:0.70    & $\pho 6.9$ & $\pm 2.3$  \\
    \hline 
  \end{tabular}
  \caption{\textit{%
      Measured differential cross sections as a function of 
      \qsq{} for $0.02<y<0.7$, 
      $1.5 < \ptds < 9\gev$ and $|\etads|<1.5$. The systematic 
      uncertainties are assumed to be the same as those for the kinematic range
      $0.02 < y < 0.85$.
    }}
  \label{tab:dxsr}
\end{table}

\clearpage 

\begin{table}[p]
\centering
\begin{tabular}{|c|cll|}
\hline 
$\qsq$ & $\siggp$ & $\Delta_{\rm stat}$ & $\Delta_{\rm syst}$ \\ 
($\gev^2$) & \multicolumn{3}{c|}{(\nb)} \\
\hline
$\sim 0$         & $729$   & $\pm 46$     &   $ ^{+110} _{-92}$\\
\hline
0.10             & $710$   & $\pm 170$    &   $ ^{+200} _{-200}$\\
0.26             & $810$   & $\pm 130$    &   $ ^{+180} _{-180}$\\
0.42             & $940$   & $\pm 200$    &   $ ^{+260} _{-260}$\\
0.59             & $890$   & $\pm 290$    &   $ ^{+370} _{-360}$\\
\hline
2.7              & $741$   & $\pm 31$     &   $ ^{+95} _{-100}$\\
7.1              & $506$   & $\pm 27$     &   $ ^{+81} _{-59}$\\
14               & $408$   & $\pm 22$     &   $ ^{+64} _{-47}$\\
28               & $278$   & $\pm 13$     &   $ ^{+36} _{-33}$\\
57               & $152$   & $\pm 13$     &   $ ^{+24} _{-24}$\\
130              & $64$    & $\pm 9$      &   $ ^{+14} _{-11}$\\
450              & $21$    & $\pm 5$      &   $ ^{+6}  _{-11}$\\
\hline 
\end{tabular}
\caption{\textit{$\gamma p$ cross sections for \dstar{} production in the
    range $1.5 < \ptds < 9\gev$ and $|\etads| < 1.5$ as a
    function of \qsq{} for $W = 160\gev$. The values at $\qsq \approx 0$ and for $\qsq
    > 2.7\gev^{2}$ are obtained from previous 
    photoproduction~\protect\cite{epj:c6:67} and DIS
    measurements~\protect\cite{pr:d69:012004} 
    in the range $1.5 < \ptds < 15\gev$ and $|\etads| < 1.5$. 
  }}
\label{tab:gammapxsect}
\end{table}

%%% Local Variables: 
%%% mode: latex
%%% TeX-master: "DESY-07-012"
%%% End: 

%% file: DESY-07-012-fig.tex
\begin{figure}[htbp]
  \begin{center}
    \includegraphics[width=\textwidth, bb=50 55 475 410]{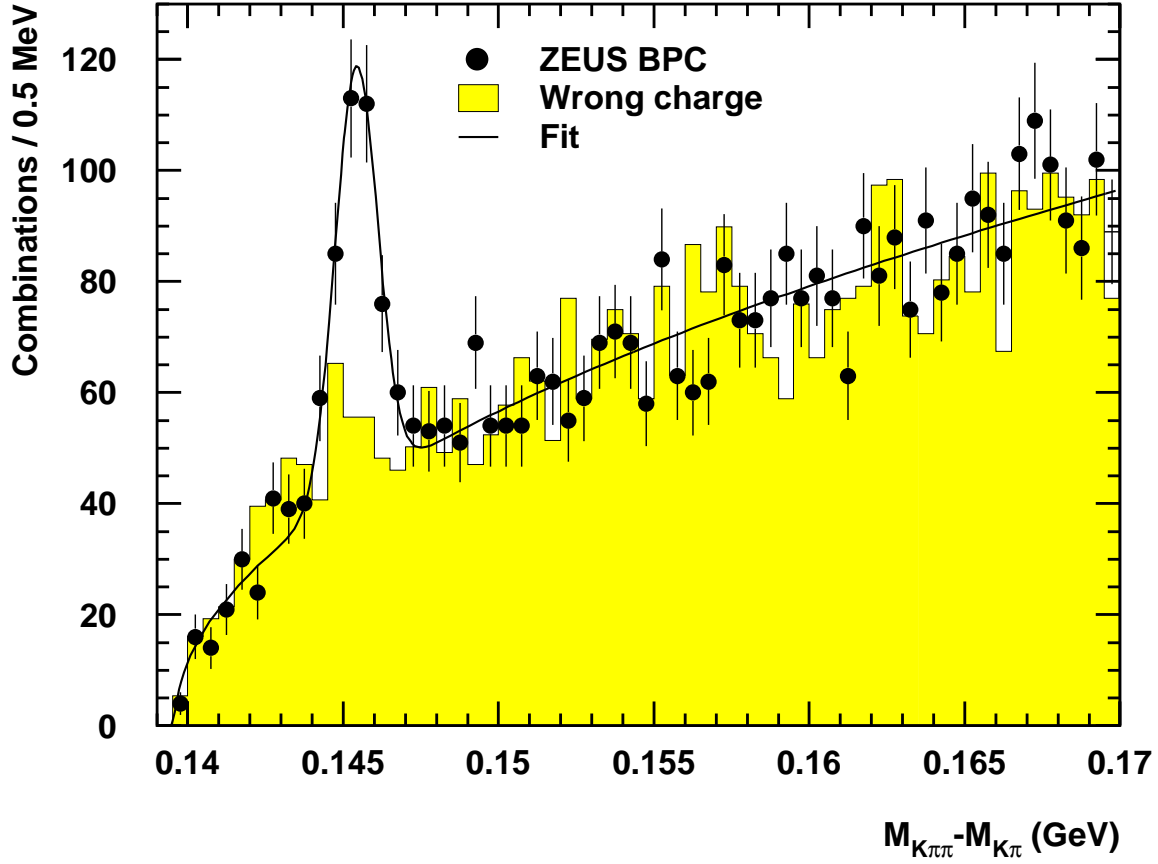}
    \caption{\textit{The distribution of the mass difference, 
      $\Delta M = M(K\pi\pi_{s})-M(K\pi)$, for \dstarpm{} candidates
      with a measured scattered electron in the BPC. 
      The histogram shows the
      $\Delta M$ distribution for wrong charge combinations, 
      normalised to the data in the region $0.151 < \Delta M <
      0.167$. The normalisation factor is 1.07.
      The solid curve is the result of the fit described in the text.
    }}
    \label{fig1}
  \end{center}
\end{figure}

\begin{figure}[htbp]
  \begin{center}
    \includegraphics[width=\textwidth, bb=25 30 530 565]{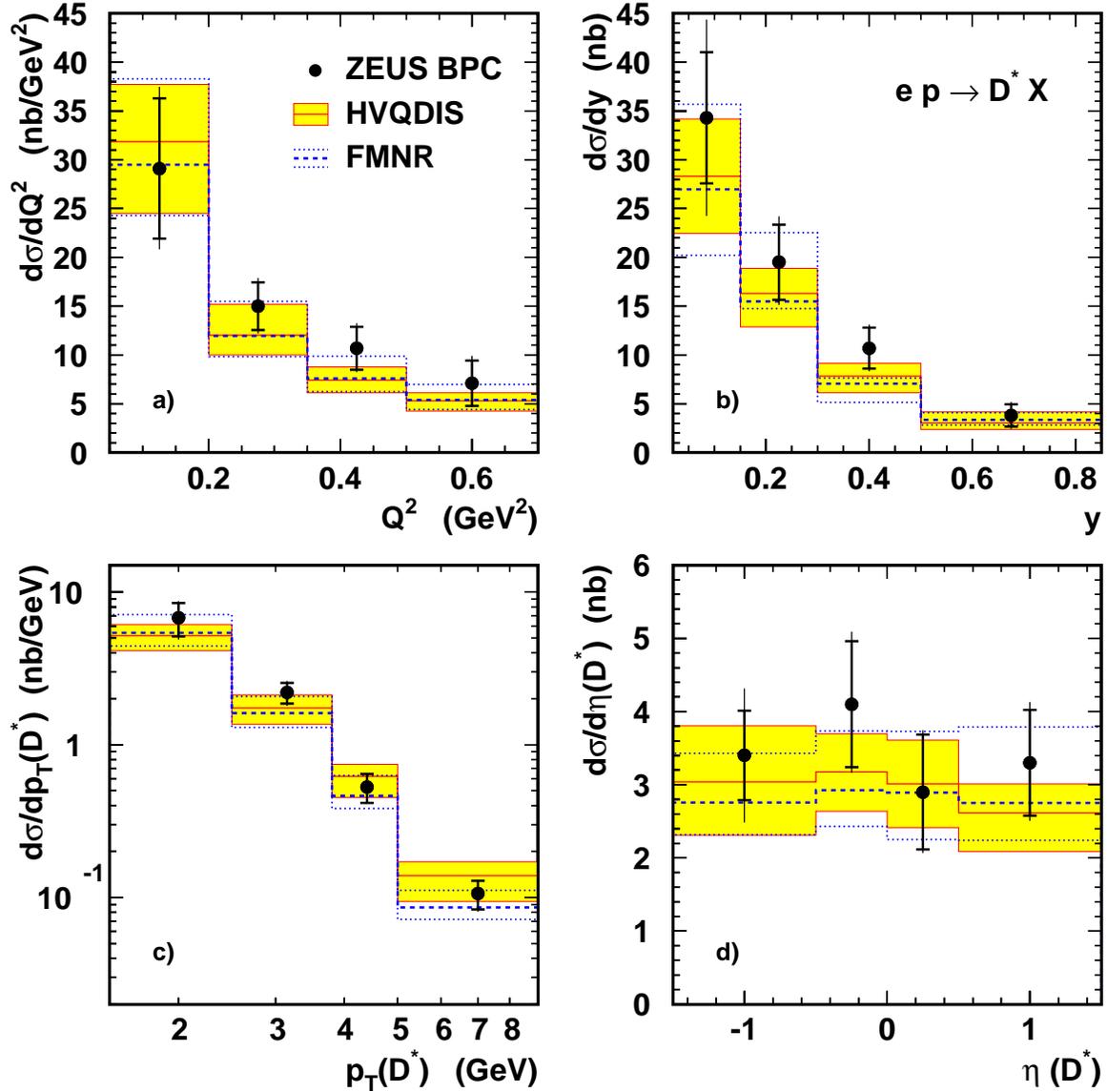}
    \caption{\textit{Differential $\dstar$ production cross sections as a function of
      (a) \qsq, (b) $y$, (c) \ptds{} and (d) \etads{} compared to the HVQDIS and
      FMNR NLO predictions.  Data are represented by points. The inner error
      bars are the statistical errors of the measurement while the open error
      bars are the sum of statistical and systematic uncertainties added in
      quadrature.  The shaded area indicates the theoretical uncertainties
      obtained by variation of the HVQDIS parameters. The dashed and dotted
      lines represent the central value of the FMNR calculation and its uncertainty,
      respectively.
    }}
    \label{fig:hvqq2yetapt}
  \end{center}
\end{figure}

\begin{figure}[htbp]
  \begin{center}
    \includegraphics[width=\textwidth, bb=45 55 475 405]{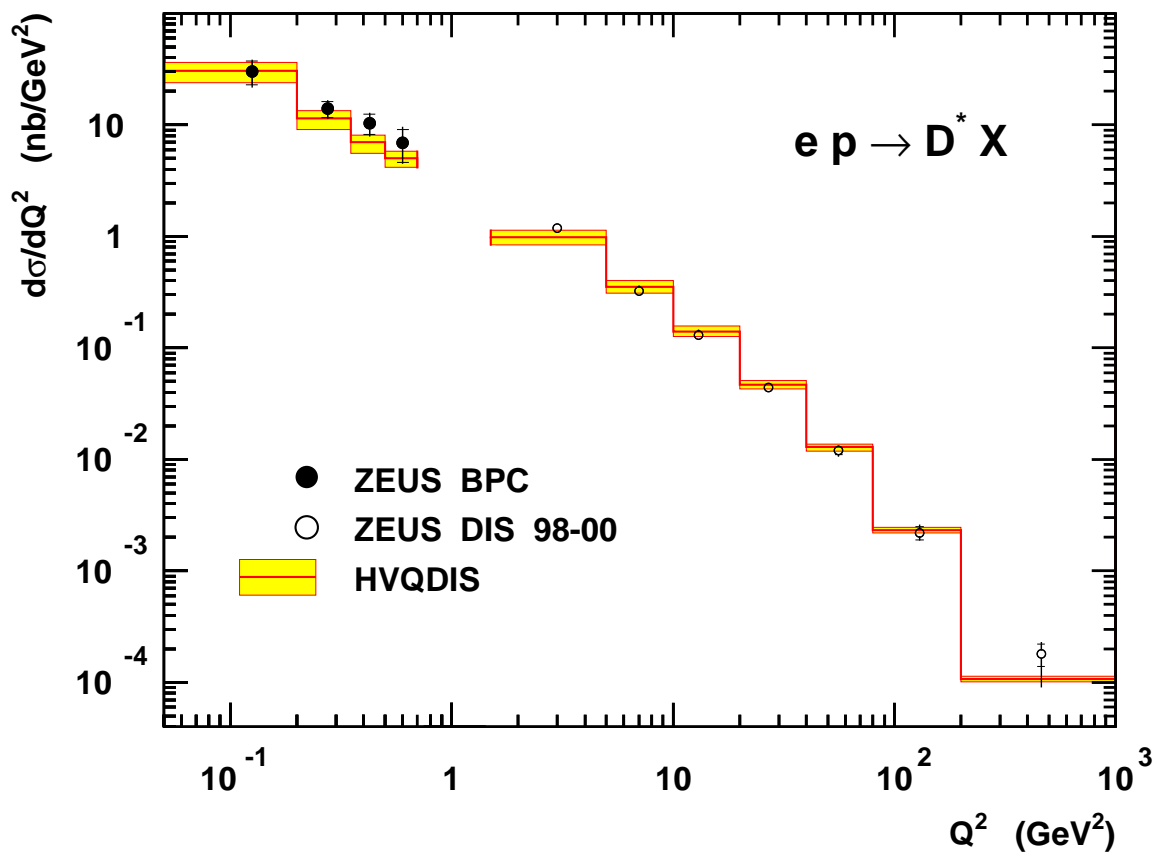}
    \caption{\textit{The \dstar{} production cross section as a function of \qsq{}
      in the kinematic region 
      $0.02 < y < 0.7$, $1.5 < \ptds < 9\gev$ and $|\etads| < 1.5$ for this
      measurement (BPC) and previous results on 
      \dstar{} production in DIS~\protect\cite{pr:d69:012004} 
      (for $1.5 < \ptds < 15\gev$), compared to the
      HVQDIS NLO prediction. The data are 
      represented by points. The inner error bars are statistical 
      while the open error bars are the sum of 
      statistical and systematic uncertainties added in quadrature.
      The shaded area indicates the theoretical uncertainties obtained by 
      variations of the HVQDIS parameters.
    }}
    \label{fig:dis2003ext}
  \end{center}
\end{figure}

\begin{figure}[htbp]
  \begin{center}
    \includegraphics[width=\textwidth, bb=45 55 535 375]{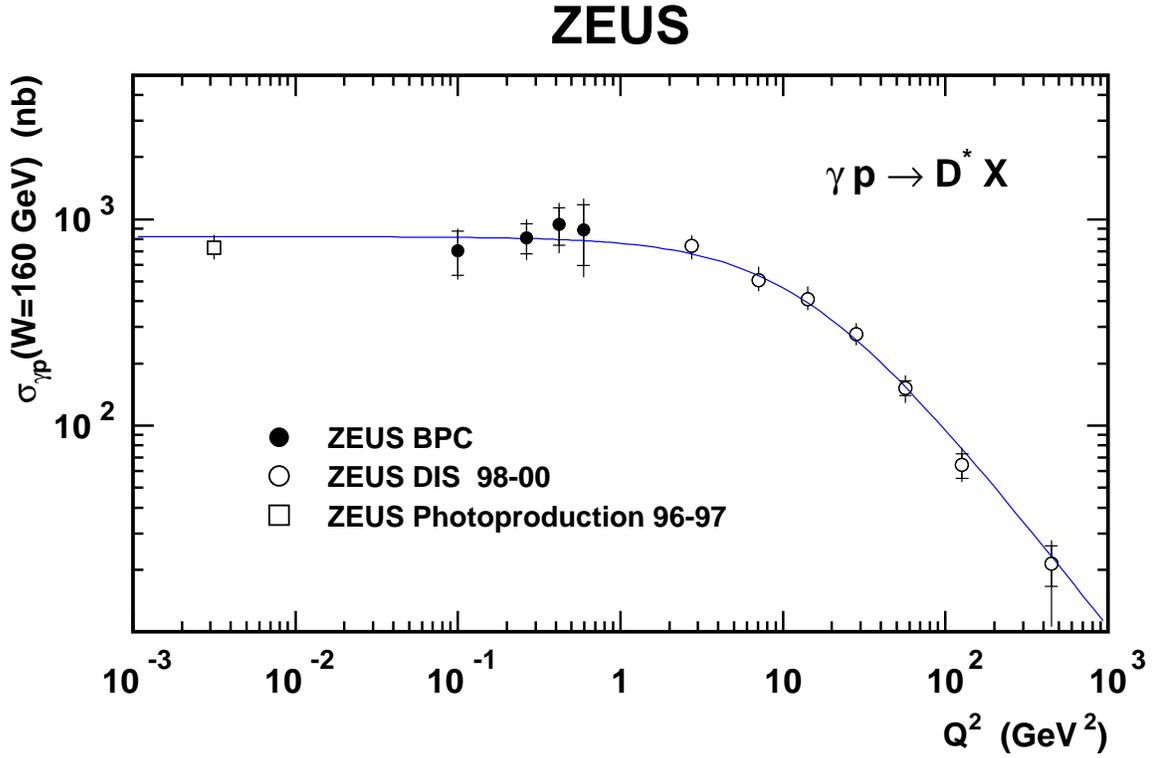}
    \caption{\textit{The $\gamma p$ cross section for \dstarpm{} production 
        in the range $1.5 < \ptds < 9\gev$ and $|\etads| < 1.5$ as a function
        of \qsq{} from this paper (BPC), compared with previous results on
        \dstar{} production in DIS~\protect\cite{pr:d69:012004} and
        photoproduction~\protect\cite{epj:c6:67} for $1.5 < \ptds < 15\gev$
        and $|\etads|<1.5$.  The data are represented by points. The inner
        error bars are statistical while the open error bars are the sum of
        statistical and systematic uncertainties added in quadrature.  The
        photoproduction point is drawn at $\qsq = 0.003\gev^{2}$ for
        convenience.  The curve shows a fit to the data described in the text.
      }}
    \label{fig:gammapxsect}
  \end{center}
\end{figure}

%%% Local Variables: 
%%% mode: latex
%%% TeX-master: "DESY-07-012"
%%% End: 

%% file: DESY-07-012-PL.bbl
\begin{mcbibliography}{10}

\bibitem{pl:b407:402}
ZEUS \coll, J.~Breitweg \etal,
\newblock Phys.\ Lett.{} B~407~(1997)~402\relax
\relax
\bibitem{np:b545:21}
H1 \coll, C.~Adloff \etal,
\newblock Nucl.\ Phys.{} B~545~(1999)~21\relax
\relax
\bibitem{epj:c12:35}
ZEUS \coll, J.~Breitweg \etal,
\newblock Eur.\ Phys.\ J.{} C~12~(2000)~35\relax
\relax
\bibitem{pl:b528:199}
H1 \coll, C.~Adloff \etal,
\newblock Phys.\ Lett.{} B~528~(2002)~199\relax
\relax
\bibitem{pr:d69:012004}
ZEUS \coll, S.~Chekanov \etal,
\newblock Phys.\ Rev.{} D~69~(2004)~012004\relax
\relax
\bibitem{np:b472:32}
H1 \coll, S.~Aid \etal,
\newblock Nucl.\ Phys.{} B~472~(1996)~32\relax
\relax
\bibitem{epj:c6:67}
ZEUS \coll, J.~Breitweg \etal,
\newblock Eur.\ Phys.\ J.{} C~6~(1999)~67\relax
\relax
\bibitem{np:b729:492}
ZEUS \coll, S.~Chekanov \etal,
\newblock Nucl.\ Phys.{} B~729~(2005)~492\relax
\relax
\bibitem{epj:c47:597}
H1 \coll, A.~Aktas \etal,
\newblock Eur.\ Phys.\ J.{} C~47~(2006)~597\relax
\relax
\bibitem{hep-ex-0608042}
H1 \coll, A.~Aktas \etal,
\newblock Preprint \mbox{hep-ex/0608042}, 2006.
\newblock Submitted to Eur.\ Phys.\ J.\ C\relax
\relax
\bibitem{pl:b407:432}
ZEUS \coll, J.~Breitweg \etal,
\newblock Phys.\ Lett.{} B~407~(1997)~432\relax
\relax
\bibitem{pl:b487:53}
ZEUS \coll, J.~Breitweg \etal,
\newblock Phys.\ Lett.{} B~487~(2000)~53\relax
\relax
\bibitem{zeus:1993:bluebook}
ZEUS \coll, U.~Holm~(ed.),
\newblock {\em The {ZEUS} Detector}.
\newblock Status Report (unpublished), DESY (1993),
\newblock available on
  \texttt{http://www-zeus.desy.de/bluebook/bluebook.html}\relax
\relax
\bibitem{nim:a279:290}
N.~Harnew \etal,
\newblock Nucl.\ Inst.\ Meth.{} A~279~(1989)~290\relax
\relax
\bibitem{npps:b32:181}
B.~Foster \etal,
\newblock Nucl.\ Phys.\ Proc.\ Suppl.{} B~32~(1993)~181\relax
\relax
\bibitem{nim:a338:254}
B.~Foster \etal,
\newblock Nucl.\ Inst.\ Meth.{} A~338~(1994)~254\relax
\relax
\bibitem{nim:a309:77}
M.~Derrick \etal,
\newblock Nucl.\ Inst.\ Meth.{} A~309~(1991)~77\relax
\relax
\bibitem{nim:a309:101}
A.~Andresen \etal,
\newblock Nucl.\ Inst.\ Meth.{} A~309~(1991)~101\relax
\relax
\bibitem{nim:a321:356}
A.~Caldwell \etal,
\newblock Nucl.\ Inst.\ Meth.{} A~321~(1992)~356\relax
\relax
\bibitem{nim:a336:23}
A.~Bernstein \etal,
\newblock Nucl.\ Inst.\ Meth.{} A~336~(1993)~23\relax
\relax
\bibitem{desy-92-066}
J.~Andruszk\'ow \etal,
\newblock Preprint \mbox{DESY-92-066}, DESY, 1992\relax
\relax
\bibitem{zfp:c63:391}
ZEUS \coll, M.~Derrick \etal,
\newblock Z.\ Phys.{} C~63~(1994)~391\relax
\relax
\bibitem{acpp:b32:2025}
J.~Andruszk\'ow \etal,
\newblock Acta Phys.\ Pol.{} B~32~(2001)~2025\relax
\relax
\bibitem{proc:chep:1992:222}
W.H.~Smith, K.~Tokushuku and L.W.~Wiggers,.
\newblock Proc.\ Computing in High-Energy Physics (CHEP), C.~Verkerk and
  W.~Wojcik~(eds.), p.~222. \newblock {Annecy, France, CERN (1992).} \newblock
  {Also in preprint \mbox{DESY 92-150B}\relax}, 1992\relax
\relax
\bibitem{thesis:tandler:2003}
J.~Tandler.
\newblock Ph.D. Thesis, Universit\"at Bonn, Bonn, Germany, Report
  \mbox{BONN-IR-2003-06 (unpublished)}, (2003),
\newblock available on
  \texttt{http://www-zeus.physik.uni-bonn.de/german/phd.html}\relax
\relax
\bibitem{nim:a515:37}
D.S.~Bailey and R.~Hall-Wilton,
\newblock Nucl.\ Inst.\ Meth.{} A~515~(2003)~37\relax
\relax
\bibitem{hep-ph-9912396}
G.~Marchesini \etal,
\newblock Preprint \mbox{Cavendish-HEP-99/17} (\mbox{hep-ph/9912396}),
  1999\relax
\relax
\bibitem{cpc:67:465}
G.~Marchesini \etal,
\newblock Comp.\ Phys.\ Comm.{} 67~(1992)~465\relax
\relax
\bibitem{cpc:86:147}
H.~Jung,
\newblock Comp.\ Phys.\ Comm.{} 86~(1995)~147\relax
\relax
\bibitem{tech:cern-dd-ee-84-1}
R.~Brun et al.,
\newblock {\em {\sc geant3}},
\newblock Technical Report CERN-DD/EE/84-1, CERN, 1987\relax
\relax
\bibitem{epj:c12:375}
CTEQ \coll, H.L.~Lai \etal,
\newblock Eur.\ Phys.\ J.{} C~12~(2000)~375\relax
\relax
\bibitem{pr:d46:1973}
M.~Gl\"uck, E.~Reya and A.~Vogt,
\newblock Phys.\ Rev.{} D~46~(1992)~1973\relax
\relax
\bibitem{jphys:g33:1}
Particle Data Group, W.-M.~Yao \etal,
\newblock J.~Phys{} G~33~(2006)~1\relax
\relax
\bibitem{thesis:irrgang:2004}
P.~Irrgang.
\newblock Ph.D. Thesis, Universit\"at Bonn, Bonn, Germany, Report
  \mbox{BONN-IR-2004-016 (unpublished)}, (2004),
\newblock available on
  \texttt{http://www-zeus.physik.uni-bonn.de/german/phd.html}\relax
\relax
\bibitem{pr:d57:2806}
B.W.~Harris and J.~Smith,
\newblock Phys.\ Rev.{} D~57~(1998)~2806\relax
\relax
\bibitem{np:b454:3}
S.~Frixione, P.~Nason and G.~Ridolfi,
\newblock Nucl.\ Phys.{} B~454~(1995)\relax
\relax
\bibitem{pl:b319:339}
S.~Frixione \etal,
\newblock Phys.\ Lett.{} B~319~(1993)~339\relax
\relax
\bibitem{pr:d67:012007}
ZEUS \coll, S.~Chekanov \etal,
\newblock Phys.\ Rev.{} D~67~(2003)~012007\relax
\relax
\bibitem{pr:d27:105}
C.~Peterson \etal,
\newblock Phys.\ Rev.{} D~27~(1983)~105\relax
\relax
\bibitem{hep-ex-9912064}
L.~Gladilin,
\newblock Preprint \mbox{hep-ex/9912064}, 1999\relax
\relax
\bibitem{epj:c44:351}
ZEUS \coll, S.~Chekanov \etal,
\newblock Eur.\ Phys.\ J.{} C~44~(2005)~351\relax
\relax
\bibitem{np:b565:245}
P.~Nason and C.~Oleari,
\newblock Nucl.\ Phys.{} B~565~(2000)~245\relax
\relax
\bibitem{prep:15:181}
V.M.~Budnev \etal,
\newblock Phys.\ Rep.{} 15C~(1974)~181\relax
\relax
\end{mcbibliography}
